\documentclass[twocolumn,showpacs,aps,prl,superscriptaddress,floatfix]{revtex4}
\usepackage{graphicx}
\usepackage{dcolumn}
\usepackage{bm}

\begin{document}

\preprint{}

\title{Magnetoelastics of a spin liquid: X-ray diffraction studies of Tb$_{2}$Ti$_{2}$O$_{7}$ in pulsed magnetic fields} 

\newcommand{\tto}{Tb$_{2}$Ti$_{2}$O$_{7}$}
\newcommand{\hto}{Ho$_{2}$Ti$_{2}$O$_{7}$}
\newcommand{\dto}{Dy$_{2}$Ti$_{2}$O$_{7}$}
\newcommand{\eto}{Er$_{2}$Ti$_{2}$O$_{7}$}
\newcommand{\yto}{Yb$_{2}$Ti$_{2}$O$_{7}$}
\newcommand{\gto}{Gd$_{2}$Ti$_{2}$O$_{7}$}

%
\author{J.P.C. Ruff}
\affiliation{Department of Physics and Astronomy, McMaster University,
Hamilton, Ontario, L8S 4M1, Canada}
\affiliation{Advanced Photon Source, Argonne National Laboratory, Argonne, Illinois 60439, USA}
\author{Z. Islam}
\affiliation{Advanced Photon Source, Argonne National Laboratory, Argonne, Illinois 60439, USA}
\author{J.P. Clancy}
\affiliation{Department of Physics and Astronomy, McMaster University,
Hamilton, Ontario, L8S 4M1, Canada}
\author{K.A. Ross}
\affiliation{Department of Physics and Astronomy, McMaster University,
Hamilton, Ontario, L8S 4M1, Canada}
\author{H. Nojiri}
\affiliation{Institute for Materials Research, Tohoku University, Katahira, Sendai 980-8577, Japan}
\author{Y.H. Matsuda}
\affiliation{Institute for Solid State Physics, University of Tokyo, Chiba 277-8581, Japan}
\author{H.A. Dabkowska}
\affiliation{Department of Physics and Astronomy, McMaster University,
Hamilton, Ontario, L8S 4M1, Canada}
\affiliation{Brockhouse Institute for Materials Research, McMaster University,
Hamilton, Ontario, L8S 4M1, Canada}
\author{A.D. Dabkowski}
\affiliation{Department of Physics and Astronomy, McMaster University,
Hamilton, Ontario, L8S 4M1, Canada}
\affiliation{Brockhouse Institute for Materials Research, McMaster University,
Hamilton, Ontario, L8S 4M1, Canada}
\author{B.D. Gaulin} 
\affiliation{Department of Physics and Astronomy, McMaster University,
Hamilton, Ontario, L8S 4M1, Canada}
\affiliation{Brockhouse Institute for Materials Research, McMaster University,
Hamilton, Ontario, L8S 4M1, Canada}
\affiliation{Canadian Institute for Advanced Research, 180 Dundas Street West, Toronto, Ontario, Canada M5G 1Z8}

\begin{abstract} 

We report high resolution single crystal x-ray diffraction measurements of the frustrated pyrochlore magnet {\tto}, collected using a novel low temperature pulsed magnet system.  This instrument allows characterization of structural degrees of freedom to temperatures as low as 4.4 K, and in applied magnetic fields as large as 30 Tesla.  We show that {\tto} manifests intriguing structural effects under the application of magnetic fields, including strongly anisotropic giant magnetostriction, a restoration of perfect pyrochlore symmetry in low magnetic fields, and ultimately a structural phase transition in high magnetic fields.   It is suggested that the magnetoelastic coupling thus revealed plays a significant role in the spin liquid physics of {\tto} at low temperatures.

\end{abstract} 
\pacs{75.80.+q, 75.10.Kt, 78.70.Ck, 75.30.Gw}

\maketitle 

Geometrically frustrated magnets have occupied the attention of physicists for decades, due to their natural inclination towards unconventional ground states \cite{diep}.  Broadly, these materials share the common trait that the interactions between magnetic ions cannot all be simultaneously satisfied, due to the connectivity of the lattice.  This gives rise to a large manifold of nearly degenerate low energy states.  The manner in which a material selects an ultimate ground state from within this manifold is generally novel, since this selection will occur due to weak perturbations commonly ignored in conventional magnets.  Hence, geometrically frustrated magnets are host to a plethora of exotic phenomena, including order-by-disorder \cite{villain,champion}, spontaneous low dimensional magnetic correlations\cite{rossyto}, three dimensional spin-Peierls-like distortions \cite{Lee00}, and even a failure to achieve long-range order at all \cite{GBrev,Gardner99}.

Among the most extensively studied families of frustrated magnets are the rare-earth-metal titanate pyrochlores \cite{MJGrev}.  These materials crystallize in the Fd$\bar3$m space group, with trivalent rare-earth ions occupying the highly frustrated pyrochlore lattice, a face-centered cubic arrangement of corner-sharing tetrahedra.  The interplay between crystal electric field (CEF) induced anisotropy, near-neighbour exchange, and long range dipolar interactions, all occurring on a lattice that can be frustrated for both ferromagnetic (FM) and antiferromagnetic (AFM) interactions, leads to widely varied behaviour depending on the rare-earth-metal ion involved.  {\hto} and {\dto} are the canonical ``spin ice'' materials \cite{GBrev}, which are frustrated ferromagnets and a magnetic analog to the famous proton disorder problem in water ice \cite{pauling}.  These materials exhibit no long range order at any measurable temperature, and exotic magnetic monopole-like excitations \cite{monopole}.  This phenomenology is understood in the context of dominant local Ising-like anisotropy and strong dipolar couplings.  {\eto} and {\yto} present the opposite anisotropy, with magnetic moments confined to local easy-planes.  Both exhibit unconventional ordered states at low temperatures, where quantum and thermal fluctuations seem to play an important role \cite{champion,ruffeto,hodges,rossyto}.  

At first glance, one might expect {\tto} to be less interesting than its cousins.  As in the spin ices, the Tb$^{3+}$ moment interacts with the surrounding CEF leading to an easy-axis anisotropy\cite{Gingras2000,mirebeaunew}.  Unlike the spin ices, susceptibility measurements give a negative Curie-Weiss constant, implying net AFM interactions\cite{Gingras2000}.  Since FM interactions are frustrated for an Ising-like pyrochlore magnet, and AFM interactions are not, it would be reasonable to assume that {\tto} should have a unique long-range ordered ground state.  However, it has been conclusively shown that {\tto} remains in a collectively paramagnetic or ``spin liquid'' state down to the lowest measurable temperatures\cite{Gardner99,Gingras2000}.  The explanation for this surprising failure to order remains elusive, although there is evidence that the relatively small anisotropy gap between the Tb$^{3+}$ ion ground state doublet and first excited state doublet is involved\cite{hamid1,hamid2}.  The anisotropy gap in {\tto} ($\Delta \sim 18$K) is comparable to the strength of the interactions between moments ($\theta_{CW} \sim -19$K)\cite{Gingras2000}, whereas the gap values in the canonical spin ices are an order of magnitude larger\cite{GBrev}.  This allows for quantum fluctuations to effectively renormalize the interactions between spins \cite{hamid1,hamid2}.  It also allows for spin orientations to effectively ``push'' against the cage of oxygen atoms surrounding the magnetic ion, opening the door to magnetostrictive effects \cite{aleksandrov,mamsurova88}.

In fact, before {\tto} was of interest to the frustrated magnetism community, the compound was primarily studied for its unusual magnetoelastic behaviour.  Over twenty years ago, researchers found a precipitous drop in Young's modulus\cite{mamsurova86}, accompanied by the onset of giant magnetostriction (MS)\cite{aleksandrov,mamsurova88} as the material is cooled below the relevant energy scale of $\sim 20$K $\sim \Delta \sim \theta_{CW}$.  These effects were explained in terms of a purely single-ion origin, involving the doublet-doublet energy level structure described above.  Although magnetoelastics were largely initially ignored when physicists became interested in frustration physics in {\tto}, there has recently been a renaissance of interest in lattice effects  on the peculiar magnetism in this compound.  It has subsequently been revealed that structural fluctuations exist within the spin liquid state\cite{ruffxray}, that applied pressure induces magnetic order\cite{mirebeaunature}, and that inversion symmetry may be violated at the Tb$^{3+}$ site at low temperatures\cite{ramantto}.

With this in mind, we have undertaken a thorough investigation of the lattice properties of {\tto} at low temperatures and in applied magnetic fields.  Single crystal samples were grown at McMaster University using the floating zone technique, with growth conditions similar to those reported previously\cite{Gardner98}.  In order to probe the lattice directly over a wide range of field and temperature, we have made use of a new 30 Tesla split-pair pulsed magnet recently commissioned at the Advanced Photon Source\cite{islamrsi}.  This system employs two closed-cycle cryostats to independently cool the sample and a small resistive split-coil of Tohoku design, which are co-aligned in the incident synchrotron x-ray beam.  The coil is connected to a capacitor bank which can repeatedly discharge kilovolt scale pulses, generating a half-sine-wave current profile with a width of $\sim$ 1 ms and a peak field up to $\sim$ 30 Tesla.  Data are collected as a function of time during the pulse using fast photon detectors.  We used an avalanche photo-diode binned to a multi-channel scalar for point-by-point data collection, and a linear array silicon strip detector for greater reciprocal space coverage.  These detectors allow for rapid read-out of data, which is binned into 10.24 $\mu$s and 20 $\mu$s frames respectively.  A full field dependence between zero and peak field can be collected in a single pulse of the magnet.  Data was collected on beamline 4ID-D at the Advanced Photon Source, with 10.885 keV photons.  The results presented in this letter comprise the first experimental study that makes use of the APS pulsed magnet system, and the seminal demonstration of its utility.  

\begin{figure} 
\centering 
\includegraphics[width=8.5cm]{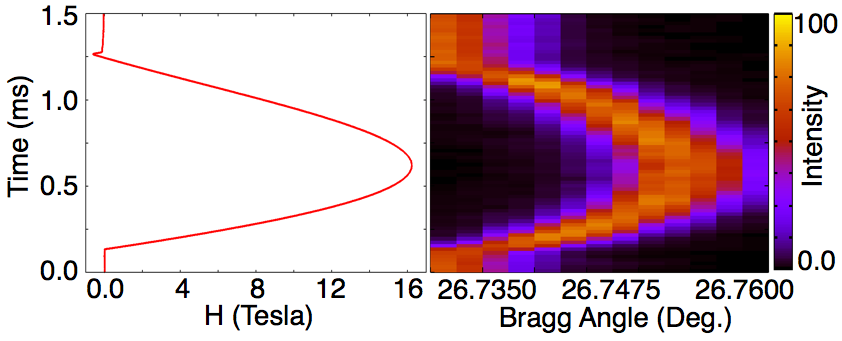}
\caption {(Colour online) Typical time-resolved diffraction data showing magnetostriction.  Left:  Magnetic field generated at the sample as a function of time, for a charging voltage of 1.1kV.  Right:  False-colour intensity map of a $\theta-2\theta$ scan of pulses through the (008) Bragg peak.  As the magnetic field increases, the peak intensity shifts to higher Bragg angles due to a contraction of the c-axis lattice parameter.}
\label{fig:1}
\end{figure}

Earlier studies of {\tto} found that the material has a giant longitudinal MS at low temperatures, wherein the crystal lattice expands along the applied magnetic field direction and contracts in the transverse direction\cite{aleksandrov}.  The magnitude of the relative change in length along the applied field direction was found to approach $(\Delta$$l/l)_{//} \sim$ $6\times10^{-4}$ at H = 6 Tesla and T = 4 Kelvin, a value at least two orders of magnitude larger than that presented by typical paramagnets and other members of the rare-earth-metal titanate family\cite{mamsurova88}.  The magnitude of the transverse contraction $(\Delta$$l/l)_{\perp}$ was found to be smaller than the longitudinal response by a factor of $\sim$ 2.5.  Due to the geometry of the scattering experiments reported here, it was necessary to apply the magnetic field transverse to the scattering plane, and therefore we measure the transverse magnetostriction (TMS) only.  A typical illustration of how this is done is shown in Fig. 1.  In this case, data is collected by scanning through the (008) Bragg peak position along the direction of momentum transfer, and pulsing at each point.  As the magnetic field increases (in this case, to a peak value of 16.2 Tesla), the intensity shifts to higher Bragg angles as the c-axis lattice parameter contracts.  As the field falls back to zero, the Bragg peak follows the field back to the original position.

\begin{figure} 
\centering 
\includegraphics[width=8.5cm]{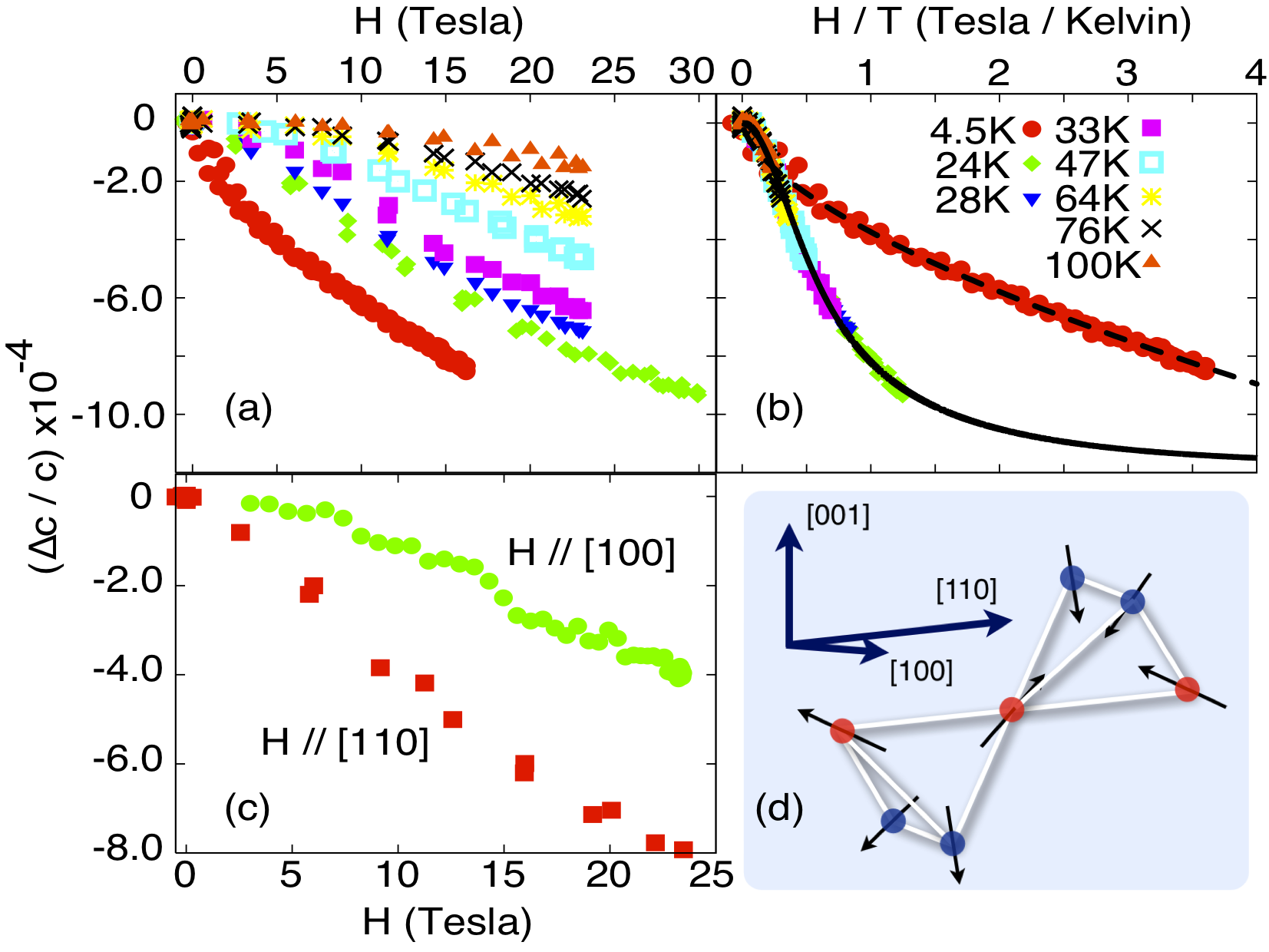}
\caption {(Colour online) TMS in {\tto}.  (a) Temperature dependence of TMS, measured on the (008) Bragg peak with H $//$ [110].  (b)  The same data, with the magnetic field rescaled by temperature.  Data with T $>$ 20 K collapse onto a single curve, well described by the square of a Brillouin function (solid line).  At lower temperatures, TMS is empirically described by a power law (dashed line).  (c)  Anisotropy of TMS (T $\sim$ 20 K).  (d) A subsection of the pyrochlore lattice.  Blue (dark) spins have easy-axes transverse to [110], while red (light) spins do not.  All spins have a component of their easy-axis along [100].}
\label{fig:2}
\end{figure}

With the demonstrated ability to measure TMS in hand, we are now able to investigate the temperature and applied magnetic field dependence of the effect.  Fig. 2 shows the results of a series of measurements similar to those demonstrated in Fig. 1.  In these plots, the relative change in c-axis lattice parameter is derived as $(\Delta$$c/c) = (\frac{\sin\theta_{B}(0)}{\sin\theta_{B}(H)} - 1)$, where $\theta_{B}(H)$ is the Bragg angle of the peak at a given magnitude of magnetic field H.  This quantity is a microscopic measurement of the TMS defined above for bulk deformations, and should be exactly analogous.  Fig. 2a shows the temperature dependence of TMS with magnetic field applied along [110].  The effect is strongly enhanced at low temperatures, and no saturation is observed up to 30 Tesla.  A more informative analysis can be made by rescaling the magnetic field axis by the temperature, as shown in Fig. 2b.  Notably, all data measured for T $> 20$ K collapses onto the same curve, which is well described by a Brillouin function squared.  This is perhaps not surprising, since a simple paramagnetic MS should be a function of the square of the magnetization\cite{engdahl}.  However, at low temperatures (within the spin liquid regime), the functional dependence changes, and is empirically well described by a power law;  $(\Delta$$c/c) \propto (H/T)^{\alpha}$, with $\alpha \sim 0.63$.  The implication is that at high temperatures, the magnetoelastic behaviour of {\tto} is that of a conventional paramagnet with a single-ion MS mechanism and a single relevant energy scale, namely (H/T).  At low temperatures, correlations between spins become important, and the single energy scale argument no longer applies.  The cooperatively paramagnetic spin liquid is also cooperatively magnetostrictive.  Clearly, this analysis would not have been possible without the availability of very high magnetic fields, since it was necessary to cover a large range in (H/T) for temperatures above 20 K to extract the Brillouin function behaviour and identify the low temperature deviation.  It should also be noted that the observed TMS for a single crystal with magnetic field applied along [110] is much larger than bulk measurements previously reported for polycrystalline samples\cite{aleksandrov}.  To explain this discrepancy, we investigated the dependence of TMS on the direction of applied magnetic field.  Fig. 2c shows  comparison between curves measured with H applied along [110] and [100] at T= 20 K.  When the magnetic field is applied along [100], the magnitude of TMS is reduced, and in agreement with previous work.  Notably, [110] is the hard axis for magnetization, while [100] is the easy axis for magnetization.  This can be understood by inspecting Fig. 2d, where a subset of spins on the pyrochlore lattice are shown pointing along the local easy-axes.  For fields applied along [110], half of the spins (labeled by blue dots) have easy-axes that are perpendicular to the magnetic field, and are therefore hard to polarize\cite{ruffprbr}.  For magnetic fields applied along [100], all spins can be polarized and the saturation magnetization should be larger.  Therefore, when the magnetic field is applied along the hard [110] direction, the tendency to polarize competes most strongly with the CEF constraint, and the necessity for magnetic moments to distort surrounding oxygen cages is maximized.  The anisotropic response of spins to applied magnetic field direction in {\tto} in particular\cite{hamid111,ruffprbr,mirebeaunew} and the rare-earth-metal titanates in general\cite{ruff05,fukazawa,clancy} has been well documented, but anisotropy in magnetoelastic effects has not been observed before.  Indeed, the original work on MS in {\tto} claimed that the effect should be isotropic, and that assumption factored largely in calculations used to explain its origin\cite{aleksandrov}.

\begin{figure} 
\centering 
\includegraphics[width=8.5cm]{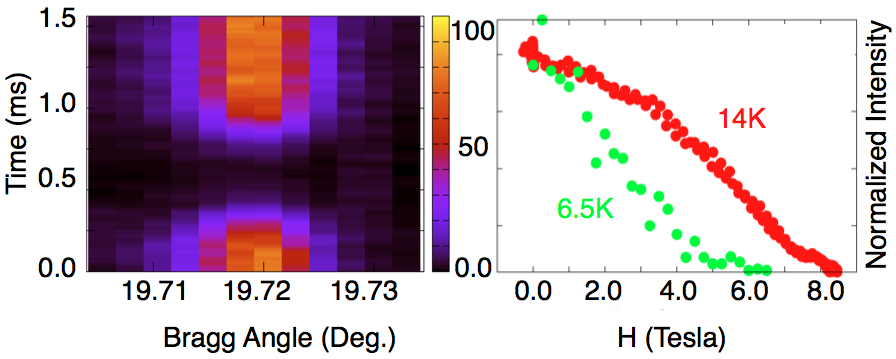}
\caption {(Colour online) Magnetic field induced suppression of (0,0,6), with H $//$ [100].  Left:  Results of a $\theta-2\theta$ scan through the peak position at 6.5K, with a peak field of 7 T (compare with Fig. 1).  Right:  Temperature dependence of integrated intensity order parameters.  (Due to high incident flux on the sample, the temperature of the scattering volume while measuring this weak feature may have been higher than the nominal values by $\sim 5$K.)}
\label{fig:3}
\end{figure}

In addition to a careful characterization of the H-T dependence of TMS, diffraction studies in pulsed magnetic fields allow for the identification of subtle deviations in the symmetries of the crystal lattice.  One known deviation in the rare-earth metal titanates is a violation of Fd$\bar3$m symmetry that leads to weak but observable peaks at $l = 2n, 2n \neq 4n$.  A perfect pyrochlore should not have Bragg peaks at (002),(006), etc.  However, features have been observed at (002) using time of flight neutron scattering in {\tto}\cite{rule}, {\hto}\cite{clancy}, and {\eto}\cite{ruffeto} that appear to be unrelated to the magnetic structure.  Here we focus on the (006) peak in {\tto}, which we found to be present up to room temperature, with an integrated intensity approximately six orders of magnitude smaller than allowed reflections along $l$.  We observed no temperature dependence down to 6 Kelvin.  Fig. 3 shows the dependence of the (006) peak on magnetic fields applied along the [100] direction.  Remarkably, instead of showing a shift in Bragg angle due to TMS, the (006) Bragg peak is smoothly suppressed as the magnetic field is increased, and smoothly reappears as the field is decreased.  We are able to map out order parameters at two temperatures, and show that the suppression occurs for relatively moderate values of the applied magnetic field, and that the critical field for the restoration of symmetry increases with temperature.  Although the specific atomic displacements involved are not known, the $l = 4n$ systematic absence generally arises from a four-fold screw axis along $c$.  Surprisingly, our results indicate that this condition is weakly violated in zero field, but that magnetostrictive strain restores this symmetry in applied field.  This seems to be a manifestation of microscopic or internal magnetostriction, which can be probed directly with x-ray diffraction but only inferred from bulk techniques such as dilatometry.  

\begin{figure} 
\centering 
\includegraphics[width=8.5cm]{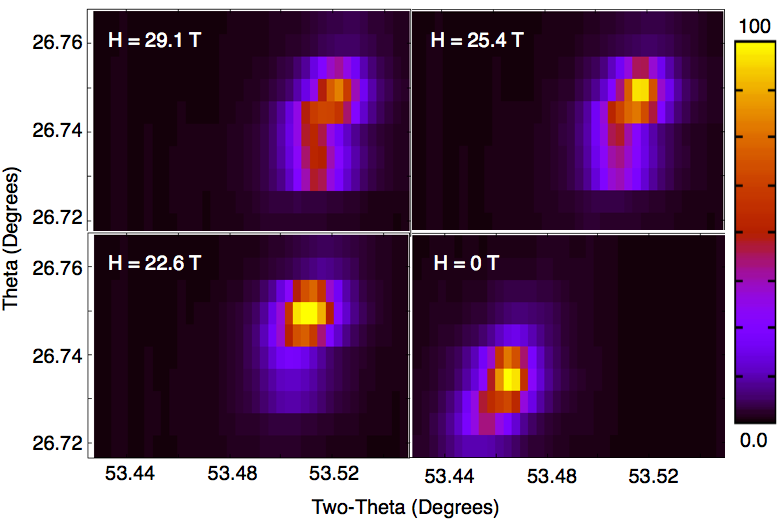}
\caption {(Colour online) $\theta$ vs. $2\theta$ mesh scans of the (008) Bragg peak, at T = 24 K and in falling magnetic field (H $//$[110]).  Above 25 Tesla, the peak is extended along the $\theta$ direction and split in $2\theta$.  Below 25 Tesla, a single peak is observed which is resolution-limited in both $\theta$ and $2\theta$.}
\label{fig:4}
\end{figure}

The low field symmetry restoration described above is a subtle effect.  We were also able to resolve a more dramatic change in lattice symmetry at higher fields.  Using a fast strip detector to collect entire $2\theta$ arcs in a single pulse, we were able to map out planes of reciprocal space by rocking the sample through the (008) Bragg peak.  Snapshots of these maps collected at T = 24 K and  for different values of magnetic field applied along [110] are shown in Fig. 4.  As the magnetic field increases during the first half of the pulse, the peak splits at a critical field of H $\sim$ 29 Tesla.  This is a hallmark of a structural phase transition with a reduction from cubic to tetragonal or orthorhombic symmetry.  Above the critical field, the peak presents a much broader profile under sample rotation ($\theta$), and two subtly distinct lattice parameters (extracted from the center of scattering in $2\theta$).  The phase transition exhibits hysteresis, and the critical falling field for a restoration of cubic symmetry appears to be H $\sim$ 25 Tesla as shown in Fig. 4.  It is important to note that the rapid rate of change of the magnetic field raises the possibility that the system is out-of-equilibrium as the field is reduced.  However, no strong hysteresis is observed in the magnetostriction under the same conditions.    The revelation of a magnetic-field-induced structural phase transition in {\tto} is a natural complement to the observation of magnetic order induced by applied pressure\cite{mirebeaunature}.  Clearly, the coupling between spin and lattice degrees of freedom is sufficiently strong that the perturbation of one changes the ground state of the other.

In conclusion, we have presented an extensive study of the magnetoelastic properties of the spin liquid material {\tto}.  This compound exhibits giant TMS, which has a strongly anisotropic dependence on the direction of the applied magnetic field.  At high temperatures, TMS is well described by a single-ion paramagnetic response, with (H/T) the only relevant energy scale.  At low temperatures (within the spin liquid state), correlations between magnetic ions modify the functional form of TMS, which is empirically described by a power law.  We have also shown that relatively weak magnetic fields applied along the crystallographic [100] direction act to restore a weakly violated symmetry of the Fd$\bar3$m space group, while large magnetic fields applied along [110] induce a symmetry-lowering structural phase transition.  As a final note, we point out that recent theoretical investigations of {\tto} imply that quantum fluctuations can effectively renormalize the interactions to be FM\cite{hamid1,hamid2}.  Since a FM ground state would necessarily create a net internal magnetic field, MS effects should couple ferromagnetism to lattice distortions even in zero applied magnetic field.  In this context, the structural fluctuations observed in {\tto} in zero field\cite{ruffxray} may be a signature of the quantum spin ice physics postulated by Molavian and coworkers\cite{hamid1,hamid2}.  We await a theoretical description of the magnetoelastic spin liquid to answer these questions, and we hope that the results presented here both motivate and constrain that effort.

Use of the APS is supported by the DOE, Office of Science, under Contract No. DE-AC02-06CH11357.  This work was supported by NSERC of Canada.


%
%






\end{document}